\documentclass[rmp,aps,twocolumn]{revtex4}
\usepackage{graphicx}

\def\bio#1#2#3#4{{\it Biophys. J. } #1;#2:#3-#4}

\def\jmb#1#2#3#4{{\it J. Mol. Biol. } #1;#2:#3-#4}
\def\jpc#1#2#3#4{{\it J. Phys. Chem. } #1;#2:#3-#4}
\def\jcp#1#2#3#4{{\it J. Chem. Phys. } #1;#2:#3-#4}

\def\pnas#1#2#3#4{{\it Proc. Natl. Acad. Sci. USA } #1;#2:#3-#4}
\def\pre#1#2#3#4{{\it Phys. Rev. E } #1;#2:#3-#4}

\def\prl#1#2#3#4{{\it Phys. Rev. Lett. } #1;#2:#3-#4}
\def\sci#1#2#3#4{{\it Science } #1;#2:#3-#4}

\begin{document}

\title{Force Unfolding Single RNAs: from Equilibrium to Far-from
Equilibrium}
\author{Fei Liu$^*$}
\author{Zhong-can Ou-Yang$^{*\dagger}$}
\affiliation{$^*$Center for Advanced Study, Tsinghua University,
Beijing 100084, China}
\affiliation{$^\dagger$Institute of
Theoretical Physics, The Chinese Academy of Sciences, P. O. Box
2735, Beijing 100080, China}

\date{\today}

\begin{abstract}
We summarize the recent simulation progress of micromanipulation
experiments on RNAs. Our work mainly consults with two important
small RNAs unfolding experiments carried out by Bustamante group.
Our results show that, in contrast to protein cases, using the
single polymer elastic theory and the well known RNA secondary
structure free energy knowledge, we can successively simulate
various behaviors of force unfolding RNAs under different
experimental setups from equilibrium to far-from equilibrium.
Particularly, our simulation would be helpful in understanding
Jarzynski's remarkable equality, which its experimental test has
received considerable attention.
\\
\end{abstract}
\maketitle

\section{Introduction}

Ribonucleic Acid (RNA) is now known to be involved in many
biological processes, such as carriers of genetic information
(messenger RNAs), simple adapters of amino acids (transfer RNAs),
and enzymes catalyzing the reactions in protein synthesis,
cleavage and synthesis of phosphodiester bonds
\cite{Cech87,Cech93}. In particular, recent discoveries indicated
that a class of RNA called small RNA operates many of cell's
control \cite[for a report, see][]{Couzin}. These diverse and
specific biological functions of RNA are guided by their unique
three-dimensional folding. Therefore, prediction or measurements
of RNA folding and folding dynamics becomes one of central
problems in biological studies.

\begin{figure}[htpb]
\begin{center}
\includegraphics[width=0.6\columnwidth]{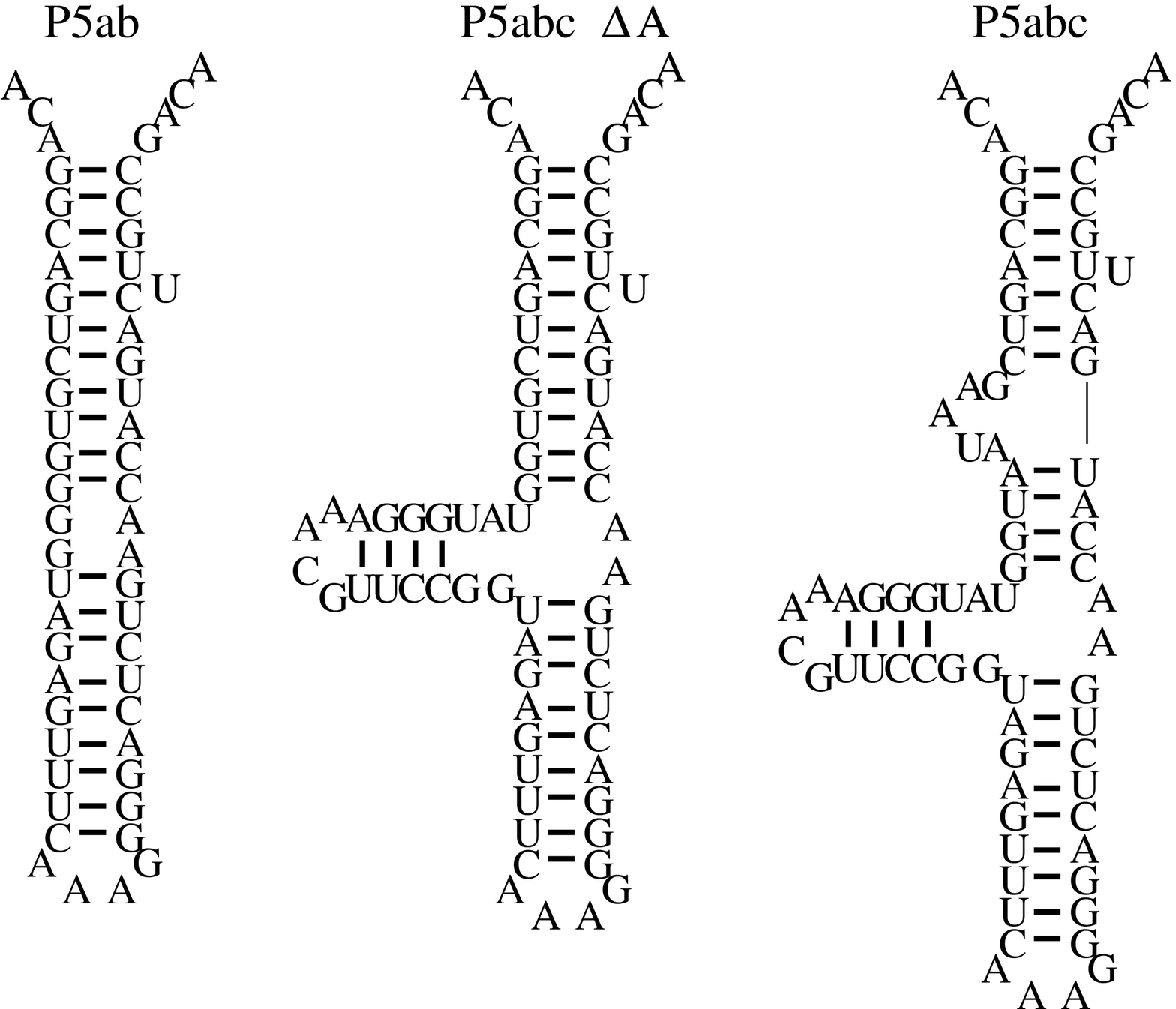}
\includegraphics[width=0.9\columnwidth]{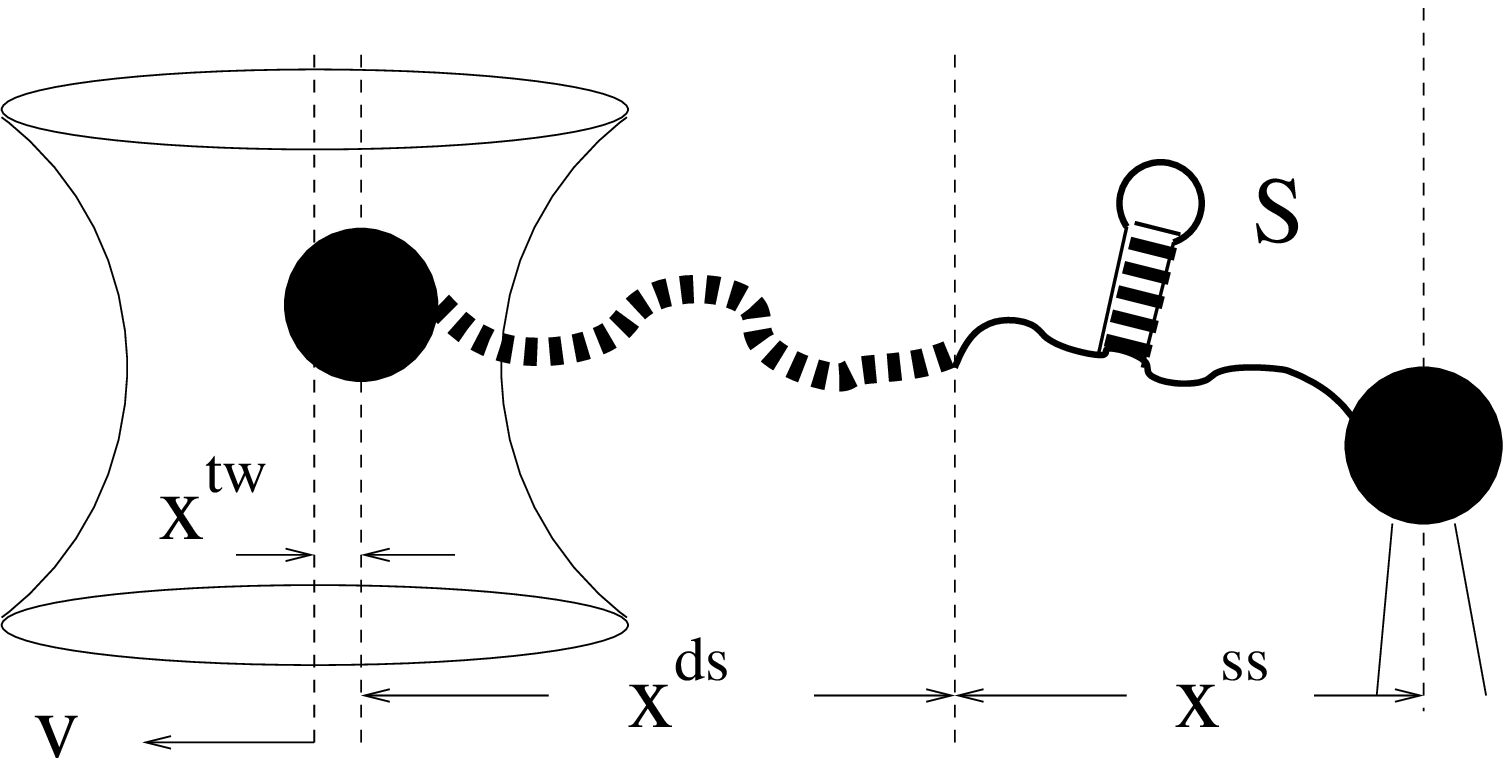}
\caption{Sketch of an optical tweezer setup and the RNA molecules
studied in the work. We denote the region between the two arcs as
the optical trap. RNA molecules are attached between the two beads
(larger black points) with a RNA:DNA hybrid handle (the black dash
curves). The center of the light trap is moved with velocity $v$.
Here the total distance at time $t$ is
$z(t)=x^{tw}+x^{ds}+x^{ss}$. The individual extensions, $x^{tw}$
the position of the bead with respect to the center of the optical
trap, $x^{ds}$ the end-to-end distance of the double-stranded DNA
(dsDNA) handles, and $x^{ss}$
 The end-to-end distance of the single RNA are freely fluctuated.
The RNA native structures for the three small RNA sequences, P5ab,
P5abc$\Delta$A, and P5abc are folded by Vienna Package 1.4.}
\label{figure1}
\end{center}
\end{figure}

In addition to standard experimental methods such as X-ray
crystallograph and NMR spectroscopy, single-molecule manipulation
technique developed in the past decade provides a fresh and
promising way in resolving the RNA folding problem
\cite{Essevaz,Harlepp,Rief,Liphardt01,Liphardt02,Onoa}. As a
concrete example, an optical tweezer setup is sketched in
Fig.~\ref{figure1} \cite{Smith,Wang}: a single RNA molecule is
attached between two beads with RNA:DNA hybrid handle; one bead is
held by a pipette, and the other is in a laser light
trap\footnote{In practice, the RNA is attached between the two
beads with two RNA:DNA hybrid handles. To simplify simulation
method, only one handle is considered. It should not change
following discussions.}. By moving the position of the pipette,
the distance between the two beads and the force acting on the
bead in the light trap can be measured with high resolution. This
sophisticated setup has showed its abilities in recording the
time-traces of the end-to-end distance of a small 22-basepair RNA
hairpin \cite{Liphardt01}, and resolving complicated unfolding
pathways of 1540-base long 16S ribosomal RNA \cite{Harlepp}.

On theoretical side, although complete three-dimensional RNA
folding prediction so far seems enormously difficult
\cite{Tinoco99}, RNA structural prediction from physical point of
view has made great progress on the level of secondary structure
\cite{Hofacker94,MacCaskill,Zuker}. The advent of the
single-molecule experiments addresses a challenging issue for
theorists: whether or how can we apply the known secondary
structural RNA knowledge to explain or predict the phenomena
observed in the single-molecule experiments? Under force
stretching or twisting, the elastic properties which were cared
little or even neglected before now must be seriously took into
account.

Many theoretical efforts have been devoted to understand force
unfolding RNAs
\cite{Montanari,Zhou1,Zhou2,Muller,Gerland,Lubensky,liuf0}.
However these theories or models are too simple to be applied in
experiments; useful free energy data about RNA secondary structure
obtained before were often neglected. Moreover, they just studied
equilibrium cases, while intriguing nonequilibrium phenomena were
beyond their scopes. Simulation method should be a good choice to
overcome these shortcomings. But we noted that, compared to
enormous simulation works about force unfolding proteins
\cite{Riefprl,Lu,Bryant,Kilmov,Socci}, the simulations for RNAs
are few \cite{Harlepp} though biological importance of the later
is the same as the former. To fill this gap, our group developed
stochastic methods and applied it to investigate the interesting
force unfolding single RNAs experiments \cite{liuf1,liuf2}. In
this paper, we will summarize our previous effort and extend them
to investigate more intriguing issue, the remarkable Jarzynski's
equality \cite{Jarzynski}, which its experimental test has
attracted consideralble attention.

\section{Model and Method}
\subsection{RNA folding without force} A
RNA sequence is denoted by a nucleotides string
$l=(x_1,x_2,...,x_n), x_i \in\{A,U,C,G\}$; the bases $x_1$ and
$x_n$ are the nucleotides at $5^\prime$ and $3^\prime$ end of the
sequence, respectively. A secondary structure $S$ of a RNA
sequence is a list of base pairs $[x_i,x_j]$ that must satisfy two
conditions: every base forms a pair with at most one other base,
and if any two base pairs $[x_i,x_j]$ and $[x_k,x_l]$ are in the
list, then $i<k<j$ implies $i<l<j$. All structures of the sequence
$l$ comprise a set, $S(l)=\{S_0,S_1,...,S_m,\bf 0\}$, here $\bf 0$
denotes the completely open chain conformation.

In order to describe the folding or unfolding process as a
time-ordered series of the structures in the set $S(l)$, a
relation $M$ which specifies whether two structures are accessible
from each other by an elementary ``move" must be reasonably
defined. The definition is identical to specifying a metric in the
set $S(l)$. Any secondary structure formation or dissolving hence
can be described by a succession of elementary steps chosen
according to some distributions from a pool of acceptable moves in
the conformational space $C(l)=\{S(l),M \}$. In the absence of
mechanical force, two kinds of move sets have been proposed in
modelling secondary structural RNA folding: one is the formation
or decay of a single helix \cite{Gultyaev,mironov}, and the other
is the removal or insertion of single base pairs per time step
\cite{Breton,flamm}. We make use of the latter, for it is the
simplest move set on the level of secondary structure. Moreover,
we mainly focus on smaller RNAs. The formation or removal of a
helix may cause larger structural changes, while its physical
relevance of RNA folding or unfolding seems debatable.

\subsection{RNA unfolding under mechanical force}
According to the difference of the external controlled parameters,
the RNA unfolding experiments can be carried out under constant
extension and constant force, i.e., the constant extension and the
constant force ensembles ~\cite{Liphardt01}.

We first consider the constant extension ensemble. Fig. ~
\ref{figure1} is the sketch of an optical tweezer setup for this
ensemble. The position of the center of the light trap is moved
according to a time-dependent relationship $z(t)=z_o+vt$, where
$z(t)$ is the distance between the centers of the light trap and
the bead held by the micropipette, $z_o$ is offset at time $t=0$
and $v$ is the constant velocity. We suppose that the changes of
the extensions of RNA and the handle proceed along one direction,
and the physical effect of the beads is negligible. Any state of
the system at time $t$ then can be specified with three
independent quantities, the extension of the RNA $x^{ss}$, the
end-to-end distance of the handle $x^{ds}$, and the RNA secondary
structure $S$, i.e., the system in $i$-state at time t
$(S_i,x^{ds}_i,x^{ss}_i)_t$. We do not include $x^{tw}$ here for
the sum of individual extensions satisfies the constraint
condition, $z(t)=x^{tw}+x^{ds}+x^{ss}$. Hence, the unfolding of
the single RNA proceeds in an conformational space
$C(l)=S(l)\times R^{ds}\times R^{ss}$, where $R^{ds}=(0,l_{ds})$
and $R^{ss}=(0,l_{ss})$ and , and $l_{ds}$ and $l_{ss}$ are the
contour lengths of the handle and the RNA molecule, respectively.
In order to describe this process as a time-ordered series of the
conformations in the space, a relation $M$ which specifies whether
two conformations are accessible from each other by an elementary
``move" (or neighbors) must be reasonably defined. We propose the
following move set \cite{liuf2},
\begin{eqnarray}
\label{moveset}
&&(S_i,x^{ds}_i,x^{ss}_i)_t\rightarrow(S_j,x^{ds}_i,x^{ss}_i)_{t'},
i\not=j\nonumber\\
&&(S_i,x^{ds}_i,x^{ss}_i)_t\rightarrow(S_i,x^{ds}_i\mp\delta,x^{ss}_i\pm\delta)_{t'},\\
&&(S_i,x^{ds}_i,x^{ss}_i)_t\rightarrow(S_i,x^{ds}_i\pm\delta,x^{ss}_i)_{t'}.
\nonumber \end{eqnarray} The first kind of the moves is the
removal or insertion of single base pairs while fixing the
extensions $x^{ds}$ and $x^{ss}$. The other two kinds are to
respectively move the positions of the end of the handle and the
end of single-stranded RNA with a small displacement $\delta$,
while the secondary structure is fixed simultaneously.

Given the system state $i$ at time t, the systematic energy can be
written as
\begin{eqnarray}
\label{systematicenergy}
E_i(t)= \Delta
G^0_i+u(x^{tw}_i)+W^{ds}(x^{ds}_i)+W^{ss}(x^{ss}_i,n_i),
\end{eqnarray}
where $\Delta G^0_i$ is the free energy obtained from folding the
RNA sequence into the secondary structure $S_i$, and the elastic
energies of the optical trap, the handle, and the single-stranded
part of the RNA are
\begin{eqnarray}
\label{energyterms}
u(x^{tw}_i)&=&\frac{1}{2}k_{tw}{x^{tw}_i(t)}^2,  \nonumber\\
W^{ds}(x^{ds}_i)&=&\int\limits_0^{x^{ds}_i}f_{ds}(x^\prime)dx^\prime,\\
W^{ss}(x^{ss}_i,n_i)&=&x^{ss}_if(x^{ss}_i,n_i)-\int\limits_0^{f(x^{ss}_i,n_i)}
x_{ss}(f^\prime,n_i) df^\prime, \nonumber
\end{eqnarray}
respectively, and $x^{tw}_i(t)=z(t)-x^{ds}_i-x^{ss}_i$. In the
expression $W^{ds}$, $f_{ds}(x^\prime)$ is the average force of
the handle at given extension $x^\prime$,
\begin{eqnarray}
f_{ds}(x^\prime)=\frac{k_BT}{P_{ds}}\left(
\frac{1}{4(1-x^\prime/l_{ds})^2}-\frac{1}{4}+\frac{x^\prime}{l_{ds}}
\right),
\end{eqnarray}
and $P_{ds}$ is the persistence length of double stranded handle.
In the expression $W^{ss}$, $x_{ss}(f^\prime,n_i)$ is the average
extension of the single stranded part of the RNA whose bases
(exterior bases) are $n_i$ at given force $f^\prime$,
\begin{eqnarray}
x_{ss}(f^\prime,n_i)=n_ib_{ss}[\coth(\frac{f^\prime
P_{ss}}{k_BT})-\frac{k_BT}{f^\prime P_{ss}}],
\end{eqnarray}
where $b_{ss}$ and $P_{ss}$ are the monomer distance and the Kuhn
length of the single-stranded RNA, respectively
\cite{Bustamante,Smith}. Note that $f(x^{ss}_i,n_i)$ is the
inverse function of $x_{ss}(f^\prime,n_i)$, and the contour length
of the RNA $l_{ss}=b_{ss}n$. The light trap here is simply assumed
to be a harmonic potential with spring constant $k_{tw}$. Hence
the loading rate is $r=k_{tw}v$.

In the real experiments, constant force can be imposed on RNA
molecules with feedback-stabilized optical tweezers capable of
maintaining a preset force by moving the beads closer or further
apart. Including the feedback mechanism in theoretical study is
not essential now. Therefore the energy of tweezer in Eq. ~
\ref{systematicenergy} is replaced by $-f\times
(x^{ss}_{i}+x^{ds}_i)$.

\subsection{Continuous time Monte Carlo algorithm}
Given the move sets and the unfolding conformational spaces, the
RNA unfolding for the two ensembles can be modelled as a Markov
process in their respective spaces. Following conventional
stochastic kinetics of chemical reactions, these processes are
described as the master equation,
\begin{eqnarray}
\label{mastereq}
\frac{dP_i(t)}{dt}=\sum_{j=0}[P_j(t)k_{ji}-P_i(t)k_{ij}],
\end{eqnarray}
where $P_i(t)$ is the probability of the system being i-state at
time t, and $k_{ij}$ is the transition probability from i-state to
j-state.

The form of the master equation looks relatively simple, however
it is mathematically intractable to solve analytically for simple
``reaction" system such as RNA P5ab. Previous work has
demonstrated that a continuous time Monte Carlo simulation is an
excellent approach toward the stochastic process described by Eq.
\ref{mastereq} \cite{flamm,BKL,Gillespie}. As a variant of the
standard Monte Carlo method, the {\it continuous time Monte Carlo}
(CTMC) method is very efficient and fast because of lacking of
waiting times due to rejection. In contrast to standard MC method,
instead of the MC step used to approximate the real time, the
``time" in Gillespie's method could be real if the transition
probabilities were calculated by first principles or empirically.

The key formula in CTMC is that, given the system at $i$-state at
current time $t$, the probability density $p(j,t'|i,t)$ that the
next state is j and it occurs at time $t'$ is
\begin{eqnarray}
\label{probability} p(j,t'|i,t)=k^{t'}_{ij}\exp(-\int_t^{t'}\sum_l
k^{\tau}_{il}d\tau),
\end{eqnarray}
where $k^{\tau}_{ik}$ are transition probabilities from the
i-state to the neighbouring $k$-state at time $\tau$, which can be
time-dependent or time-independent (no parameters $\tau$)
\cite{Gillespie,JansenA}, and the sum is over all neighbors of
$i$-state. According to Eq. \ref{probability}, then the time $t'$
for the next state to occur can be obtained by solving the
following equation,
\begin{eqnarray}
r_1=\exp(-\int_t^{t'}\sum_l k^{\tau}_{il}d\tau),
\end{eqnarray}
where $r_1$ is a uniform random number in the interval [0,1]. For
time-independent situation, the equation reduces to most common
expression $r_1=\exp[-(t'-t)\sum_l k_{il}]$. While if the
transition probabilities involve time $t$, then numerical methods
for integration and root finding have to be applied \cite{Press}.
Then the next state $j$ is chosen if another uniform random number
$r_2\le\sum_{l=1}^{j}k^{t'}_{il}/\sum_l k^{t'}_{il}$.

We assume that the transition probabilities satisfy the symmetric
rule \cite{Kawasaki}
\begin{eqnarray}
k^{t}_{ij}=\tau_o^{-1}\exp(-\beta (E_j(t)-E_i(t))/2),
\end{eqnarray}
where $\tau_o$ scales the time axis of the unfolding process from
the experimental measurements. Apparently, the transition
probabilities satisfy the detailed balance condition.

\subsection{Partition function method in equilibrium}
If the moving velocity of the light trap vanishes, an exact
partition function method can calculate the molecular average
extension and the average force under the given distance $z$
\cite{Gerland}. The method is an extension of the partition
function method proposed for RNA secondary structural prediction
\cite{MacCaskill}. Different from the experimental measurements of
the free energy with slow pulling velocity (quasi-equilibrium
process) \cite{Liphardt02}, we obtain the equilibrium information
by this exact method. Considering coincidences of formulae and new
physical quantities needed, we rewrite the formulae in
Ref.~\cite{Gerland}.

The key idea of the exact method is that the partition function
over all secondary structures of a given RNA can be calculated by
dynamic programming \cite{MacCaskill}. Given the partition
function $Q(i,j,m)$ on the sequence segment [i,j] with exterior
bases $m$, its recursion formula is as follows,
\begin{eqnarray}
Q(i,j,m)&=&{\bf 1} \delta_{m,j-i+1}+qb(i,\Delta+j-m) \nonumber\\
&&+\sum_{k=i}^{j-1}\sum_{l=1}^{k-i+1}
Q(i,k,l)\\
&&\times qb(k+1,l+\Delta+j-m),\nonumber
\end{eqnarray}
where $\Delta=2$, the partition function $qb(i,j)$ on the sequence
segment [i,j] for which the $i$ and $j$ bases are paired; Vienna
Package 1.4 provides the calculation codes \cite{Hofacker}.

For the constant extension ensemble, let the partition function of
the RNA molecule at extension $x$ (including the handle) be
$Z_n(x)$. Then the function can be written as
\begin{eqnarray}
Z_n(x)&=&\sum_m^n \int_0^{l_{ds}}\int_0^{mb_{ss}}dx^{ds}dx^{ss}
\delta
(x-x^{ds}-x^{ss})\nonumber \\
&&Q(1,n,m)\exp(-\beta W(x^{ds},x^{ss},m))
\end{eqnarray}
where $W(x^{ds},x^{ss},n)= W^{ds}(x^{ds})+W^{ss}(x^{ss},n)$. The
molecular free energy landscape along $x$ then is
$G_o(x)=-\beta^{-1}\ln Z_n(x)$. To calculate the average force
$\langle f\rangle$ and the average extension $\langle x\rangle$ at
given distance $z$, the systematic partition function ${{\cal
Z}_n}(z)$ required is
\begin{eqnarray}
{{\cal Z}_n}(z)=\int_0^z dx Z_n(x)\exp(-\beta u(z-x)).
\end{eqnarray}
Then the systematic free energy $G(z)=-\beta^{-1}\ln {{\cal
Z}_n}$, $\langle f\rangle=\partial G(z)/\partial z$ and $\langle
x\rangle=z- \langle f\rangle /k_{tw}$.

While for the constant force ensemble, the partition function
${\cal Z}_n(f)$ under given force $f$ is
\begin{eqnarray}
{{\cal Z}_n}(f)=\int_0^{(l_{ds}+l_{ss})} dx Z_n(x)\exp(\beta f x),
\end{eqnarray}
and the average extension $\langle x\rangle=\beta^{-1}\partial \ln
{{\cal Z}_n}/\partial f$.

\subsection{Parameters and measurement}
We simulate the single RNA folding and unfolding under mechanical
force at the experimental temperature $T=298 K$. The elastic
parameters used are: $P_{ds}=53$ nm, $l_{ds}=320$ nm,
$b_{ss}=0.56$ nm, $P_{ss}=1.5$ nm, and $k_{tw}=0.2$ pN/nm. We use
the single-stranded DNA parameters for the single stranded part of
RNA because they have similar chemical structure. The displacement
$\delta=0.1$ nm. The free energy parameters for the RNA secondary
structures are from the Vienna package 1.4 \cite{Hofacker} in
standard salt concentrations $[Na^+]=1$ M and $[Mg^{2+}]=0$ M. In
addition to the standard Watson-Crick base pairs (AU and CG), GU
base pair is allowed in our simulation. Formation of an isolated
base pairs is forbidden because of their instability. In the
constant extension ensembles, the force $f_i$ acting on the RNA
molecule at i-state is calculated by $f_i=k_{tw}x^{tw}_i$, and the
bead-to-bead distance $x^{bb}_i=x^{ds}_i+x^{ss}_i$. In the
constant force ensemble, the extension of the molecule is just
$x^{bb}_i$.

\section{Results and discussion}
\subsection{Single RNAs thermodynamics}
\begin{figure}[htpb]
\begin{center}
\includegraphics[width=0.9\columnwidth]{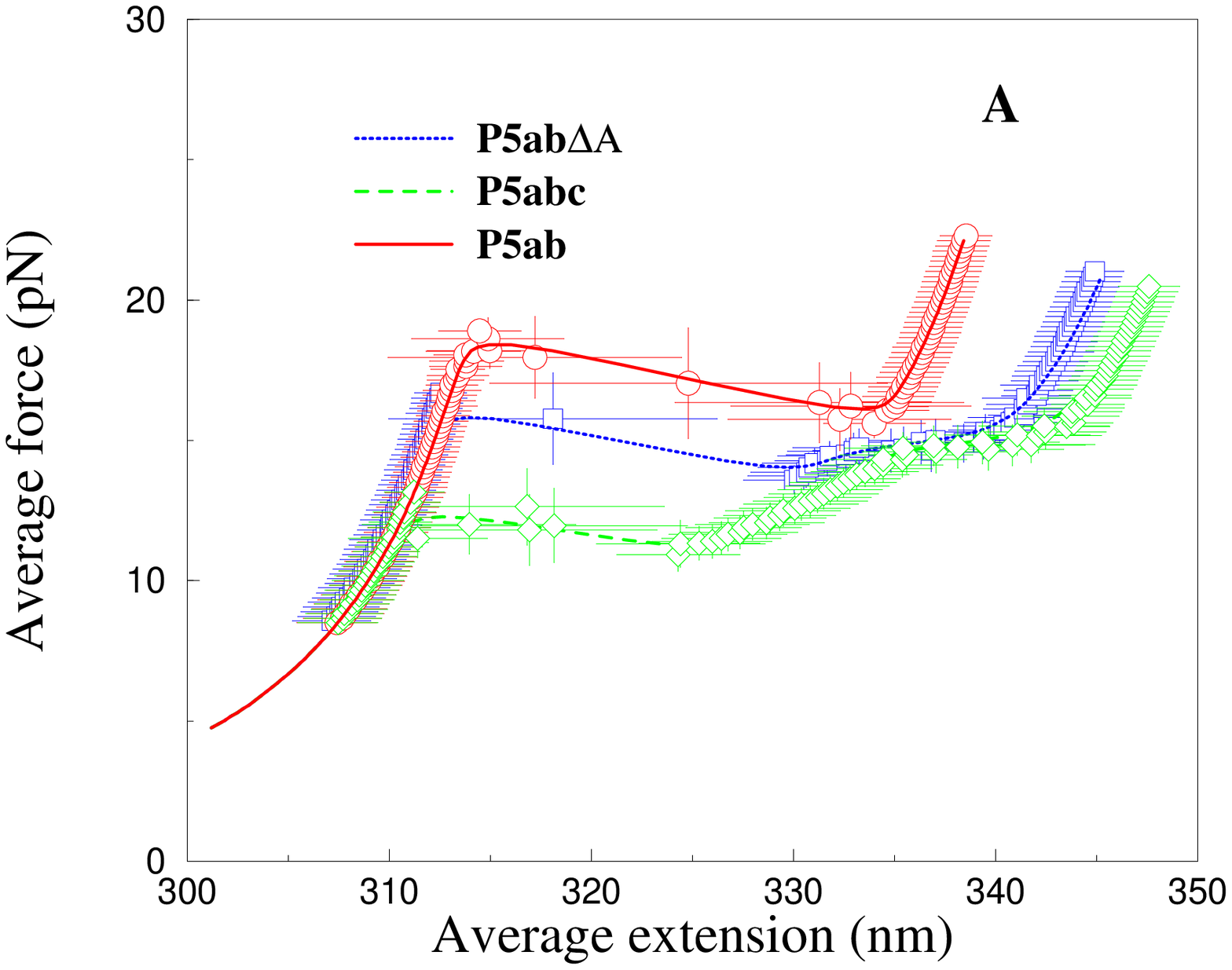}\\
\includegraphics[width=0.9\columnwidth]{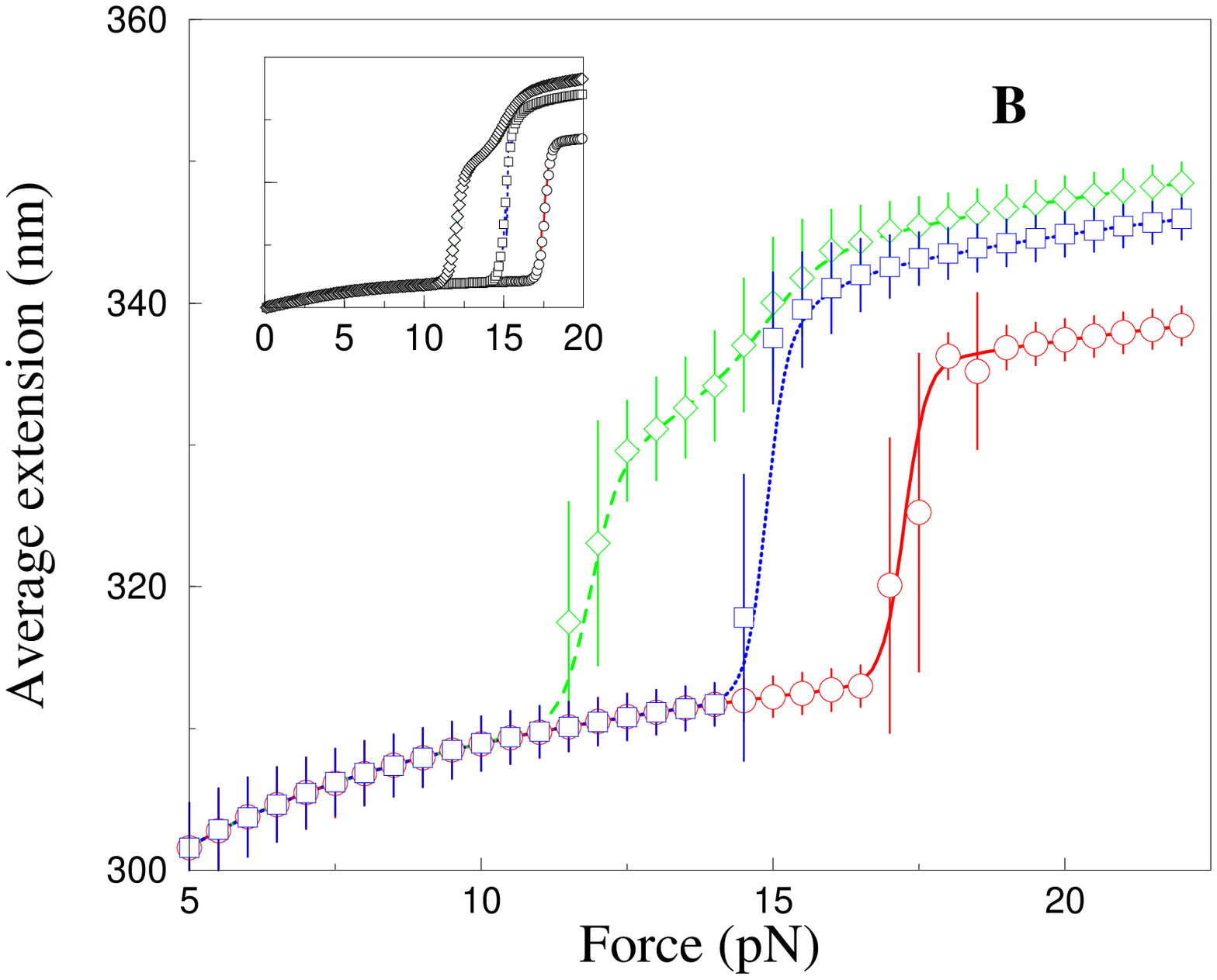}\\
\caption{Comparison of the exact and simulation force-extension
curves in equilibrium for P5ab, P5abc$\Delta$A and P5abc on the
constant extension (A) and the constant force (B) ensembles. The
different symbols are from the simulation methods, and the
different lines are from the exact methods. They agree with each
other very well. Inset, force-extension curves for the same
ensemble recalculated by another move set. } \label{figure2}
\end{center}
\end{figure}

A comparison between our simulation in equilibrium and the
prediction of the exact partition function method should be
helpful in confirming the correctness of our method. We simulate
the average force-extension curves of the three RNA molecules for
the two ensembles with standard approach: the average physical
quantity $A$ is calculated according to $\langle
A\rangle=\tau^{-1}\int_{0}^\tau A(t)dt$, here $\tau=10^6$; see
Fig. ~\ref{figure2}. The force-extension curves calculated by the
exact method are plotted with different kinds of curves. We can
find that these two independent calculations agree very well.

Although the two methods quite consist each other, the values of
the unfolding forces have apparent discrepancies with experimental
measurements. For example, in the absence of $Mg^{2+}$ the values
are 13.3, 11.3 and 8.0 pN for P5ab, P5abc$\Delta$A, and P5abc
molecules for the constant extension ensemble, respectively
\cite{Liphardt01}. It is not strange because we do not include the
effect of ionic concentration in our model. Hence we choose a
reasonable ionic correction of RNA free energies \cite{cocco}.
Unfortunately, we still do not get good results; see Table I.
There are tow possible causes leading to such discrepancies. One
is that the ionic corrections or free energy parameters for RNA
are not precise enough; they cannot be used in the force unfolding
cases. The other is that polymer elastic parameters are not very
precise. We prefer to the later. In addition to RNA free energy
measured and tested for almost forty years, the persistent length
of ssDNA in ionic environment is still debatable \cite{Murphy}.
For instance, we calculate the unfolding forces of the three
molecule with $P_{ss}=2.2$ nm and indeed find that they are closer
to the experimental values. As a further demonstration, we also
list other values measured in previous experiments and compare
them with theoretical predictions in the same table.

\begin{table*} \caption{The unfolding forces $f_u$ of
different molecules under different experimental conditions. The
experimental data are from the previously published data
\cite{Liphardt01,Rief,Bustamanterev}. The theoretical values are
from the exact numerical methods developed above, where $f^i_u$,
$i=1,2,3$ represent the unfolding forces without the ionic
correction, with the ionic correction on the free energy and with
the ionic and the persistent length corrections, respectively.
Here We do not show the P5abc unfolding force for it is not
reversible in $Mg^{2+}$ due to the presence of tertiary
interactions.}
\begin{center}
\begin{tabular}{cccccccc}
\hline\hline Molecule& temperature (K) &  $Na^+$ (mM) & $Mg^{2+}$
(mM)& $f^1_u$ (pN)& $f^2_u$ (pN)& $f^3_u$ (pN) &
$f^{exp}_u$ (pN)\\
\hline
P5abc&298&250&0&12.2&11.4&10.0&7.0-11.0\\
poly(dA-dU)&293&150&0&12.3&11.0&9.3&9.0 \\
P5abc$\Delta$A&298&250&0&15.8&14.8&13.2&$11.4\pm0.5$\\
P5abc$\Delta$A&298&250&10&&15.4&13.8&$12.7\pm0.3$\\
P5ab &298&250&0&18.4&17.4&15.7&$13.3\pm1.0$\\
P5ab &298&250&10&&18.0&16.2&$14.5\pm1.0$\\
CG hairpin&293&150&0&25.8&24.4&22.4&17.0\\
poly(dC-dG)&293&150&0&25.1&23.8&21.7&20.0\\
\hline\hline
\end{tabular}
\label{table}
\end{center}
\end{table*}

\subsection{Single RNAs kinetics}
\subsubsection{Constant extension ensemble}
\begin{figure}[htpb]
\begin{center}
\includegraphics[width=0.9\columnwidth]{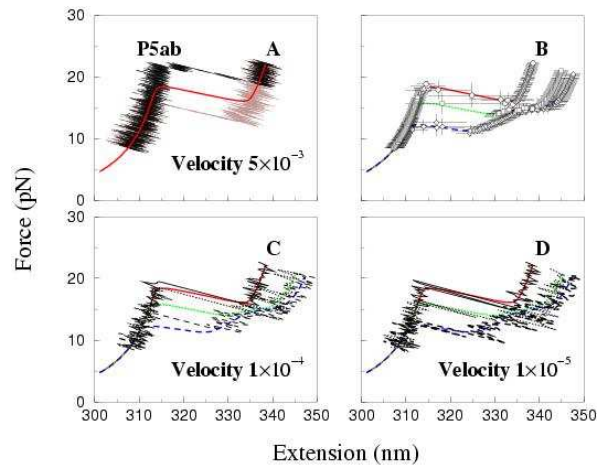}
\caption{ A. One of the time trajectories of unfolding and
refolding for P5ab with velocity $5\times10^{-3}\AA$.
Force-hysteresis is observed. B. Fig. 2A is showed here again. C.
The unfolding force-extension curves recorded at unit time $10^5$
for the molecules with velocity $1\times 10^{-4}\AA$. D. The
force-extension curves recorded at time unit $10^6$ with velocity
$1\times 10^{-5}\AA$.} \label{figure3}
\end{center}
\end{figure}

{\bf Force-extension curves.} As an example, we stretch P5ab
molecule with the velocity $v=5\times 10^{-3}\AA$ from the offset
$z_o=350$ nm to $z(t)=450$ nm, and then relax it with the same
velocity; here we let $\tau_o=1$. One of the time trajectories is
showed in Fig. \ref{figure3}A. Apparently, the unfolding and
refolding trajectories are not coincident, i.e., force-hysteresis,
which indicates that the molecule is driven from thermodynamic
equilibrium \cite{Liphardt01} occurs.

In experiments record of the force and extension at given times
with a slow velocity is a more common method in the equilibrium
measurement. Hence we simulate the three curves with the two slow
velocities $1\times10^{-4}\AA$ and $1\times10^{-5}\AA$. Because
enormous data would be generated if the time trajectories were
recorded, we only show the data per unit times $10^5$ and $10^6$
(see Figs. ~\ref{figure3}C and D). For the faster velocity, we
find that, except P5ab case, the unfolding forces for the others
do not equal the equilibrium values; whereas for the later, the
curves of simulations consist with the exact curves. It means that
the unfolding of the three molecules with $1\times10^{-5}\AA$ is
or near equilibrium. Two features in Fig. \ref{figure3}D are of
our interest. Compared with the curves obtained by the time
averaging, the curves recorded at time points are very rough even
before and after the unfolding. And although the whole extension
$z(t)$ monotonically increases with time, the extensions of the
molecules may still jump between two values, such as P5ab and
P5abc molecules. Indeed similar phenomena were also observed in
the experiment \cite{Liphardt01}. They indicate the fluctuations
of the extension and RNA structures under the force. Just because
of this observation, and that the phenomena are not rare in
simulation and the experiment.

In the experiment \cite{Liphardt01,Liphardt02}, the unfolding
P5abc are near-equilibrium and far from equilibrium at the loading
rates 2-5 pN/s and 34-52 pN/s, respectively (similar values for
P5abc$\Delta$A). And our simulations also show that the unfolding
the same molecule are near-equilibrium and far from equilibrium at
the velocities $10^{-5}\AA$ and $10^{-4}\AA$, respectively. Let
them be equal correspondingly we then can estimate the constant
$\tau_o\approx 10^{-7}$s. We will scale the time with this
parameter below for convenience.

\begin{figure}[htpb]
\begin{center}
\includegraphics[width=0.9\columnwidth]{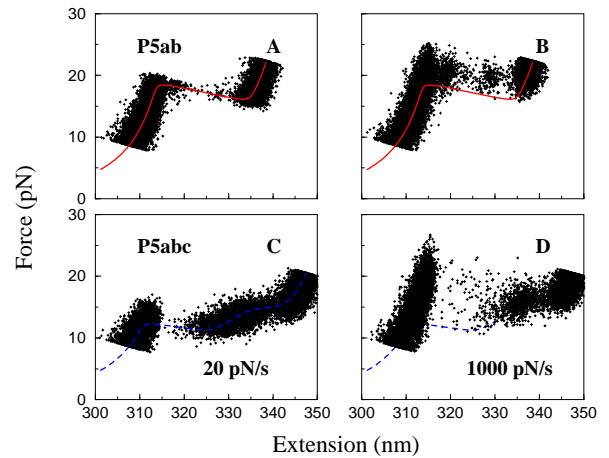}
\caption{ Unfolding trajectories of P5ab (A,B) and P5abc (C,D)
with loading rates 20 and 1000 pN/s. Curves (superposition of 100
curves per figure) are represented by 100 points with the equal
time interval; for clarity we do not connect them with lines.}
\label{figure4}
\end{center}
\end{figure}

Fig. ~\ref{figure4} shows 100 trajectories with two loading rates
for P5ab and P5abc molecules. The trajectories are stretched from
the same offset $z_o=350$ nm after thermal equilibrium until the
terminal extension $z=450$ nm. For both the loading rates, below
and above the unfolding forces, the force-extension curves are
dominated by the double-stranded handles. But the values of the
unfolding forces apparently fluctuate and dependent on the rate
and the molecular type. When the pulling speed is faster, or the
loading rate is larger (1000 pN/s), the average unfolding force
increases correspondingly. This phenomenon has been theoretically
predicted earlier \cite{Evans}. We note that at the same loading
rate, the trajectories of P5ab are closer to its equilibrium
force-extension curve than the trajectories of P5abc. It means
that the relaxing process of the former is faster than the latter.
This fact has also been observed in the experiment
\cite{Liphardt01}.

{\bf Free energy reconstruction.} According to Ref. \cite{Hummer},
the unperturbed molecular free energy landscape $G_o(x)$ along the
molecular extension $x$ can be calculated from
position-versus-time curves with the expression
\begin{eqnarray}
\label{equation14}
G_o(x)-G(t=0)=\\  \nonumber -\beta
^{-1}\log\langle\delta(x-x(t))\exp (\Delta w_t)\rangle
\end{eqnarray}
where $\Delta w_t=w_t-k_{tw}(x(t)-vt)^2/2$, $G(t=0)$ is the free
energy of the whole system in equilibrium at initial time $t=0$,
and
\begin{eqnarray}
\label{equation15} w_t=k_{tw}v(vt^2/2+z_ot-\int_o^tx(t')dt')
\end{eqnarray}
The free energy landscape can be reconstructed by one time slice
according to Eq. \ref{equation14}. But considering that for finite
stretching trajectories, we only sample a small window around the
molecular equilibrium position at the whole extension $z(t)$,
Therefore, a weighted histogram method was proposed \cite{Hummer},
\begin{eqnarray}
\label{equation16} G_o(x)-G(t=0)=\\ \nonumber -\beta ^{-1}\log
\frac{ \sum_{t_i}\frac{\langle\delta (x-x_t)\exp{(-\beta
w_{t_i})}\rangle}{\langle\exp{(-\beta w_{t_i})}\rangle}}{\sum
_{t_i}\frac{\exp[-\beta u(x(t_i),t_i)]}{\langle\exp(-\beta
w_{t_i})\rangle}},
\end{eqnarray}
where the sum is over many time slices $t'$, and the average is
over the repeated trajectories at each given time slice. For each
trajectory, we choose the discrete time $t_i=i\Delta t$,
$i=1,\cdots, 100$, here $\Delta t=10/v$, i.e., the time moving the
light trap 1 nm (or every point in Fig. ~\ref{figure4}).
\begin{figure}[htpb]
\begin{center}
\includegraphics[width=0.9\columnwidth]{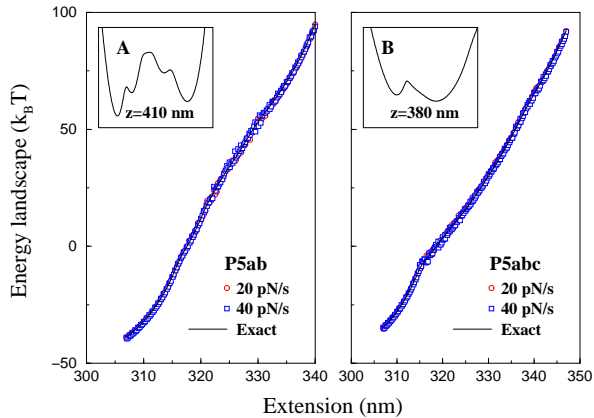}
\caption{ Comparison of the free energy landscapes of the two
molecules P5ab and P5abc reconstructed from the Jarzynski equality
with two loading rates 20 and 40 pN/s and the exact landscapes
calculated from the partition function method. The number of
trajectories for each case is 1000. The insets are the free energy
landscapes of the system composed of the molecules and the light
trap potential, which are from partition function method. Note
that we do not show the scales of the extensions and free
energies.} \label{figure5}
\end{center}
\end{figure}

Fig. ~\ref{figure5} shows the finally reconstructed free energy
landscapes for the two molecules at two loading rates 20 and 40
pN/s. The precisions of reconstructions are satisfactory. We note
that landscapes are unexpectedly trivial: neither of them presents
energy barrier. Ref. \cite{Ritort} has investigated Jarzynski's
equality by modelling RNA molecules as a two-level system with an
intermediate barrier. Our calculations apparently contradict their
assumption. Indeed, the strong unfolding-refolding cooperativity
observed in the experiments \cite{Liphardt01,Liphardt02} arises
from the coupling of the RNA molecules and the light trap; the
addition of their potentials is a two-level system (see the
respective insets in the figure). Therefore, the two-level system,
although is a good approximation in RNA folding study, should not
be simply copied to the force unfolding cases.

{\bf  Free Energy Difference Estimators} Although Jarzynski's
equality has many applications, our understanding of its behavior
is still rather limited \cite{Gore}. For example, we are not clear
whether other free energy estimators are better than the equality,
and whether the number of repeated trajectories in landscape
reconstruction above is enough or excessive to achieve reasonable
precisions. Recently such discussions have attracted considerable
interests \cite{Hummerb,Gore,Zuckerman}. Our simulation here
provides a good opportunity to numerically investigate these
questions.

Instead of the molecular free energies, we will use the systematic
free energies (the molecule and the optical trap) for simplicity.
Hence we define Jarzynski estimator $\Delta G_{JE}(z) = - \beta ^{
- 1}\log\langle\exp(-\beta w_z)\rangle_N$ (we here replace time
$t$ by the whole extension $z$ because of the linear relation
between them) and $\Delta G_{JE}(z)=G_{JE}(z)-G_{JE}(z_o)$. Two
other common estimators that can be used to calculate the free
energy differences are the mean work estimator $\Delta
G_{MW}(z)=\langle w_z\rangle_N$ and the fluctuation-dissipation
theorem estimator $\Delta G_{FD}(z)=\langle w_z\rangle_N-\beta
\sigma_w^2$, where $\sigma_w$ is the standard deviation of the
work distribution \cite{Hermans}.

\begin{figure}[htpb]
\begin{center}
\includegraphics[width=0.9\columnwidth]{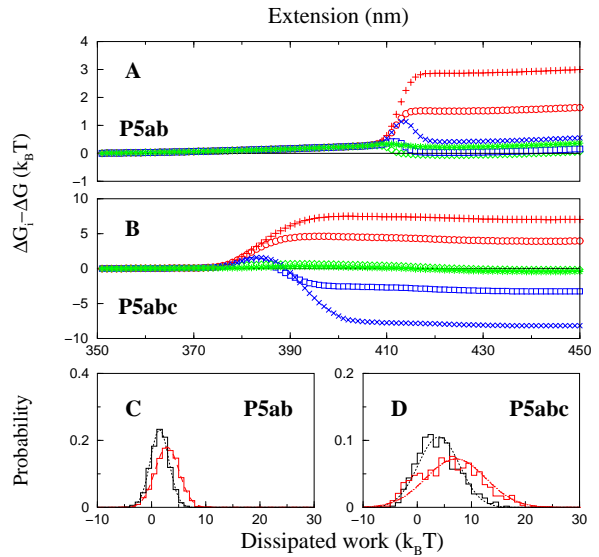}
\caption{  A and B. The differences between the three free energy
estimators and the exact energies for P5ab and P5abc at two
loading rates 20 (closed symbols) and 40 pN/s(crossed symbols): MW
circle and plus, FD square and times, and JE diamond and star.
Here N=1000. C and D. Histograms of the dissipated works at
extension $z=430$nm for P5ab and P5abc molecules at the two
loading rates. The lines are Gaussian functions with mean and
variance from the same data, where the dotted lines are for 20
pN/s, and the dashed lines are for 40 pN/s. } \label{figure6}
\end{center}
\end{figure}

To get an intuitive observations about the three estimators, we
calculate the free energy differences between the estimators and
the exact energies, $\Delta G_i(z)-\Delta G(z)$, $i=$MW, FD, and
JE. The differences of P5ab and P5abc with the loading rates 20
and 40 pN/s respectively are showed in Figs. ~\ref{figure6}A and
B; here we choose N=1000.

There are two common features in the figure. First is that the
free energy differences for each estimator are not uniform along
the distance. For example for JE estimator, the differences are
maximum around the unfolding distances such as 415 nm for P5ab.
Therefore we conclude that nonequilibrium behaviors of the same
molecule are not uniform along its extension, even if the RNA is
unfolded with the same loading rate. Then for both the molecules,
the JE estimator is always better than the MW at any loading
rates. For the P5ab case, the FD estimator is more or less better
than the JE as the extension $z>415$ nm. This trend is more
apparent as P5ab is unfolding with smaller loading rate 20 pN/s.
In contrast, the JE estimator for P5abc is superior to the FD
estimator over the entire extension range at the two loading
rates.

Because the above analysis is carried out at a given $N$-value, we
do not consider the errors caused by different samples. In
addition, we also do not consider how the results depend on the
number of trajectories. Ref. \cite{Gore} has studied the issues in
two extreme situations: the large N limit and the system in the
near-equilibrium regime. Because of $N\le 100$ in the current real
experiment, in the present work we investigate them in the middle
value of N, although it is not hard to run $10^5$ trajectories in
our simulation ``experiment". Previous analysis has found that
different from the case far from equilibrium, the properties of
the three estimators in the near-equilibrium regime are
independent of concrete models. Hence the investigation should
provide a good opportunity to test the correctness of the
unfolding kinetics we designed.

\begin{figure}[htpb]
\begin{center}
\includegraphics[width=0.9\columnwidth]{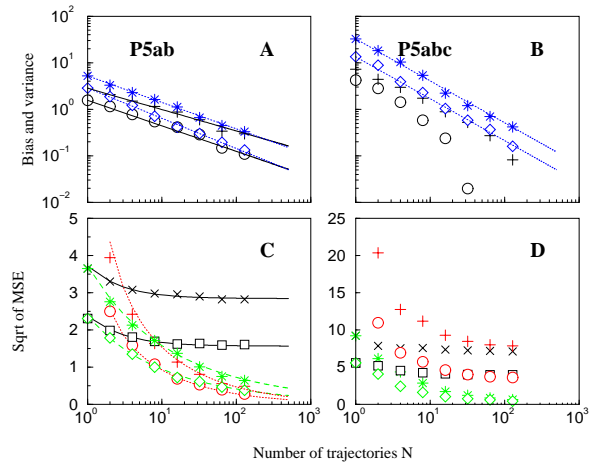}
\caption{A and B. Biases and variances dependence on the number of
trajectories for P5ab and P5abc molecules at two loading rates 20
(closed) and 40 pN/s (crossed) at the extension $z=430$ nm. C and
D. Square roots of the MSEs for the three estimators for the two
molecules at the same extensions: MW estimator square, times and
solid line; FD estimator circle, plus and dotted line; and JE
estimator diamond, star and dashed line.} \label{figure7}
\end{center}
\end{figure}

Ref. \cite{Gore} has introduced three important properties
associated with the estimators: bias $B_i(z)=\langle \Delta
G_i(z)-\Delta G(z)\rangle$, which represents systematic error
created by the finite N; variance $\sigma_i^2(z)=\langle (\Delta
G_i(z)-\langle \Delta G_i(z)\rangle)^2\rangle$ accounting for
statistical error because of different samples, and mean square
error (MSE), a standard measure for the quality of an estimator,
$MSE_i(z)=\sigma_i^2(z)+B_i^2(z)$, where i=MW, FD and JE. Although
these quantities depend on the distance $z$, we choose a
representation at z=430 nm for simplifying the discussion below:
when the extension $z$ is greater than the unfolding distance, the
qualitative behaviors of the quantities are almost the same. The
biases and variances dependence on $N$ for the JE estimators for
P5ab and P5abc are plotted in Fig. ~\ref{figure7}A and B, and the
MSEs dependence on N for the three estimators are in Fig.
~\ref{figure7}C and D, where two loading rates are 20 and 40 pN/s,
respectively, and each symbol is the average of 100 sets.

For P5ab case, we note that both the biases and variances of the
JE estimator at the two loading rates can be well approximated
with power functions. For example, at the loading rate 20 pN/s,
the bias $B_{JE}(430)=\langle
w_{dis}\rangle(430)/N^\alpha=1.57/N^{0.55}$, and $\sigma_{JE}
^2(430)=\sigma^2_w/N^\gamma=2.86/N^{0.65}$, where the dissipated
work $w_{dis}(z=430)$ is defined as $w_z-\Delta G(z)$, and
$\sigma^2_w=\langle (w_z-\langle w_z\rangle)^2 \rangle$; they also
are the bias and variance of the JE estimator with $N=1$. This
interesting observation can be understood from the distributions
of the dissipated works. Fig. ~\ref{figure6}C shows their
histograms. We find that they both agree well with Gaussian
functions whose means and variances are obtained from the
corresponding data. Indeed, according to the force-extension
curves, we argued that the unfolding of P5ab at loading rate 20
pN/s is near-equilibrium; see Fig. ~\ref{figure3}A. The good
agreement between the histograms and Gaussian function therefore
is not unexpected: when a system is in the near-equilibrium
regime, it always have a Gaussian dissipated work distribution,
and in particular, an important equality holds,
$\sigma_w=2\beta^{-1} \langle w_{diss}\rangle$ \cite{Hermans};
here $2.86\approx2\times 1.57$. Another demonstration of P5ab in
the near-equilibrium regime is that the MSEs of the three
estimators obtained by our simulations consist with the following
expressions \cite{Hummerb,Gore}:
\begin{eqnarray}
\label{equation17} MSE_{MW}&=&2\frac{\langle w_{dis}\rangle}{\beta
N}+\langle
w_{dis}\rangle^2 \nonumber \\
MSE_{FD}&=&2\frac{\langle w_{dis}\rangle}{\beta N}+2\frac{\langle
w_{dis}\rangle^2}{N-1} \\
MSE_{JE}&=&2\frac{\langle w_{dis}\rangle}{\beta
N^\gamma}+\frac{\langle w_{dis}\rangle^2}{N^{2\alpha}}\nonumber;
\end{eqnarray}
see the lines in Fig. ~\ref{figure7}C. We firstly see that MW
estimator is the worst among the three estimators. Although it is
known that the FD estimator in near-equilibrium regime is
unbiased, for smaller N, here about $\le 10$, the JE estimator is
still superior to FD. This result is from the error in estimating
$\sigma_w^2$ from limited data. When N is larger, FD estimator is
the best among them. Our observations are coincident with the
previous analysis \cite{Gore}. In addition, we also see that for
the case of 40 pN/s, larger number of trajectories will be
required to get the same quality of any estimator.

So how about the estimators far from equilibrium? The same
properties are showed in Fig. ~\ref{figure7}B and D. We see that
at the loading rates we presented, although the variances can
still be approximated with power functions, e.g.,  $\sigma_{JE}
^2(430) =13.46/N^{0.9}$ for P5abc at 20 pN/s, no such functions
present in the biases. The discrepancies of the histogram and the
Gaussian function of P5abc in Fig. ~\ref{figure6}D also reflect
it. Even if no analytic results have been obtained except the
large N limit \cite{Zuckerman,Gore}, our simulations still give
some hints about the properties of the three estimators in the
far-from-equilibrium regime: JE estimator should be the best among
the estimators; while FD and MW estimators be almost equally poor
in this regime.

\subsubsection{Constant force ensemble}
Compared to the general thermodynamics of RNA under force in
equilibrium, single-molecule methods are more interesting in
kinetic folding and unfolding studies. With the single molecule
experiments we can follow the actual folding or unfolding
trajectories of a single molecules on high resolution even when
they occur in equilibrium state, which will shed light on the
difficult kinetic folding problem \cite{liuf1,liuf2}. We mentioned
above that the extension of P5ab in the constant extension
ensemble may hop back and forth between two states. To investigate
this bistability, Ref. \cite{Liphardt01} imposed a constant force
on P5ab \cite{Liphardt01}. They found that, when the force was
held constant at the transition within $\sim 1$ pN, P5ab switched
back and forth with time from the folded hairpin (hp) to the
unfolded single strand (ss). A two-state kinetics was proposed to
explain the intriguing phenomenon. The rate constants for
unfolding reaction can be fit to an Arrhenius-like expression of
the form $k_{u}(f)\propto\exp(f\Delta x^\ddag_{u}/k_BT)$, where
$\Delta x^\ddag_{u}$ is the thermally averaged distance between
the hairpin state and the transition state along the direction of
force. A similar expression also holds for the folding rate
$k_{f}(f)$. Apparently, this description can not clarify the
physics underlying the folding and unfolding reactions.

Because our simulation is based on the microscopic interactions,
we are interested in whether the similar time traces can be
obtained by simulation. We are not ready to choose the direct move
sets in Eq. \ref{moveset}, instead another reasonable move set was
proposed \cite{liuf1} to enhance simulation efficiency. Because
what we concern about is kinetic behaviors of single RNA, for
convenience the contribution of double-stranded handles is
neglected. Indeed, under constant force the handle can be viewed
as one part of the feedback mechanism. If we model single-stranded
part of RNA structure $S_i$ as an extensible freely joined chain,
the elastic free energy contribution of it under force $f$ then is
$W^{ss}(n_i,f)=n_ik_BTb_ss/P_ss\ln \left[\sinh(u)/u\right ]$,
where $u=l_{ss} f/k_BT$. Therefore Eq. ~\ref{systematicenergy} is
largely simplified in such ensemble as
\begin{eqnarray}
\Delta G^0_i - W^{ss}(n_i,f).
\end{eqnarray}
We see that the extensional variables are absent from systematic
energy. Note the extension we record in simulation now is
$x_{ss}(f,n_i)$. Such simplification requires that under
mechanical force the conformational relaxation process of the
single stranded part of RNA is faster than the slowest process of
the secondary structural arrangement. Our discussion shows that in
contrast to the constant extension ensemble, the RNA secondary
structure $S$ alone can completely specify any state of the
constant force ensemble. Therefore, the move set is the same with
the set for RNA folding without force, i.e., the unfolding space
is $C(l)$.

we recalculate the force-extension curves for the three RNA
molecules in equilibrium; see Fig. ~\ref{figure2} inset. Except
the regions before transitions where the elastic property of the
handle dominates, the shapes of the curves and the values of
unfolding force obtained by the two different simulation methods
are almost the same.

\begin{figure}[htpb]
\begin{tabular}{cc}
\includegraphics[width=0.9\columnwidth]{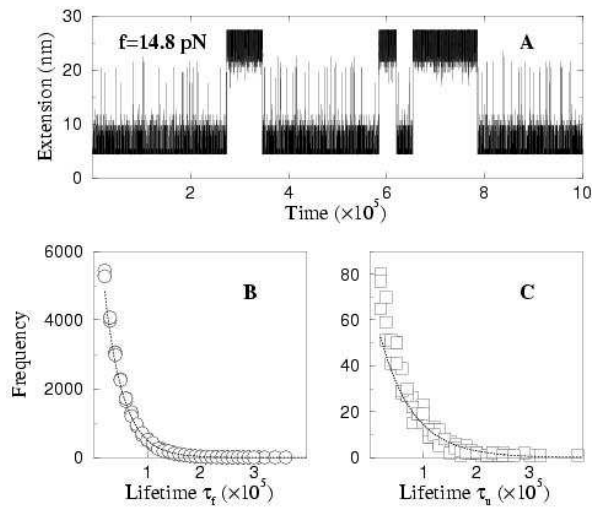}\\
\end{tabular}
\caption{Simulation for RNA p5ab kinetics. A. Extension versus
time traces of the molecule at a force in equilibrium, here the
unit of time is $\tau_o$. Frequency distributions of the lifetimes
of the single stranded (B) and hairpin states (C) at 14.2 pN. The
average lifetimes of these two states in this simulation can be
obtained by fitting to exponential functions; see the curves
therein.} \label{figure8}
\end{figure}

\begin{figure}[htpb]
\begin{center}
\includegraphics[width=0.9\columnwidth]{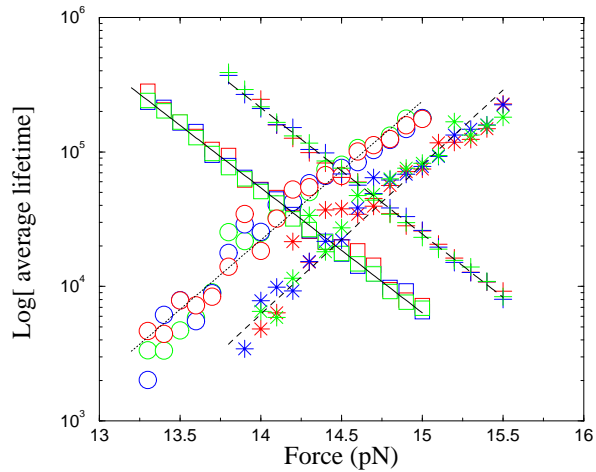}
\caption{Logarithm of the average lifetimes of single stranded and
hairpin states for p5ab molecule at difference forces around the
unfolding. The time is in unit $\tau_o$, which can be obtained by
fitting to experimental data. Note the slopes of $\ln\langle
\tau_f\rangle$ and $\ln\langle \tau_u\rangle$ are independent of
the fitting value $\tau_o$. The crossed symbols are for the
presence of $Mg^{2+}$.} \label{figure9}
\end{center}
\end{figure}

We record the extension-time traces of the RNA molecules at
different constant forces in equilibrium. In order to compare with
real data, ionic correction are took into account (see the third
corrections in Table I). For example, one extension-time traces at
force 14.8 pN for P5ab without $Mg^{2+}$ is shown in Fig.
~\ref{figure8}{A}.
The extension of the molecule jumps between two values, $\sim 5$
nm and $\sim 22$ nm around the unfolding force. Because the jumps
are extremely rapid with respect to the lifetimes of the molecule
in the two states, we simply classify the states whose extensions
are larger than 15 nm as the single stranded states, and the
others as the hairpin states. In addition, there are significant
fluctuations about the two states. Around the unfolding the
frequencies of the different lifetimes at the single stranded
state and the hairpin state can be obtained by a long time
simulation (in order to get enough data, the simulation time is
$10^9 \tau_0 $ after equilibrium). Fig. ~\ref{figure4}{c} shows
the frequency distributions of a typical simulation at force 14.2
pN. These distributions can be fit to an exponential function
$\propto \exp(-t/\langle\tau_i\rangle)$ very well, where
$\langle\tau_i\rangle, i=u,f$ denote the force-dependent average
lifetimes at the two states, respectively. For example, the
average lifetimes in this simulation are
$\langle\tau_u\rangle\approx6.2\times 10^4\tau_o$ and
$\langle\tau_f\rangle\approx3.5\times 10^4\tau_o$. We calculate
all average lifetimes near the unfolding force of P5ab, and their
corresponding values with different forces are shown in Fig.
~\ref{figure9}. We find that the logarithms of the lifetimes for
the two states are strikingly consistent with linear functions of
the forces. Because the reaction rate constants are the inverse of
the average lifetimes, we fit $\tau_o$ by making
$\langle\tau_u\rangle(f^\star)=\langle\tau_f\rangle(f^\star)$
equal to the experimental value $1/k^\star$, where $k^\star\equiv
k_u=k_f$, and had $\tau_o^{-1}=2.6\times10^5$ sec$^{-1}$. Using
the same method, the reaction rate constants for P5ab in the
presence of $Mg^{2+}$ are also calculated. A comparison of the
simulation results and the experimental data is listed in Table
II. Because our simulation does not need additional fitting
parameters, the striking consistence of our results with the
experiment assures us that the RNA folding and unfolding model
proposed here has grabbed the main physics.
\begin{table*}
\caption{Simulation results for P5ab compared to the experimental
data from Ref. \cite{Liphardt01} (in bold). }
\begin{center}
\begin{tabular}{ccccc}
\hline\hline Molecule& $\langle\Delta x\rangle (nm)$ & $f^\star$ (pN) & $\ln k_f(f)(s^{-1})$& $\ln k_u(f)(s^{-1})$\\
\hline\bf P5ab, $Mg^{+2}$ &$19\pm2$ & $14.5\pm1$ & $41\pm1.9-(2.8\pm0.1)f$ & $-39\pm2.3+(2.9\pm0.2)f$\\
P5ab, by Cocco et al.& & 15.1& $ 27.5-2.74 f$ &$-42.9+ 1.93 f$\\
P5ab, by us&20.0 & 14.7& $ 39.4-2.6 f$ &$-30.1+ 2.2 f$\\
\bf P5ab, EDTA & $18\pm2$ &$13.3\pm1$ &$37\pm4.0-(2.7\pm0.3)f$ &$-32\pm4.8+(2.6\pm0.4)f$\\
P5ab, by us &20.0 &14.2 &$35.7 -2.4f$ & $-28.3 + 2.1f$ \\
\hline\hline
\end{tabular}
\label{table}
\end{center}
\end{table*}

\section{Discussion and conclusion}

Compared to the enormous kinetic simulations of the force
unfolding proteins, the effort contributed to RNA is relatively
little. To fit the gap, we developed a kinetic stochastic
simulation to the force unfolding single RNAs. Different from
previous force unfolding method, for the constant extension
ensemble external time-dependent force can be taken into account
correctly. We make use of the algorithm to investigate the
experiment testing the important Jarzynski's equality
\cite{Liphardt02}. Instead of the force versus extension
trajectories used in the experiment, the extension versus time
trajectories are used to reconstruct the free energy landscapes,
compare the qualities of the three estimators, and investigate the
estimators dependence on the trajectories number. For the constant
force ensemble, we particularly studied the interesting
extension-time traces dependence on force, which would be relevant
to RNA folding dynamics.

The most advantage in study of force unfolding RNAs is that the
knowledge accumulated in the past forty years for RNA secondary
structure provides a solid fundament for theoretical predictions
(including kinetics and thermodynamics) in practice. Therefore, we
believe that our model would be useful in RNA biophysical studies
in the future. Of course several improvements still can be added
in our algorithm, e.g., adding the effects of $Mg^{+2}$ to include
complicated tertiary interactions \cite{Onoa}. Recent works have
shown that the inclusion of pseudoknots is possible
\cite{Isambert,rivas}.\\

The computation of this work was performed on the HP-SC45 sigma-X
parallel computer of ITP and ICTS, CAS. We thank Dr. Flamm for
providing us his computer program KINFOLD. F.L. thanks Dr. F.Ye,
R.-L. Dai, and Y. Zhang for their supporting in the computation.
This research was supported by the theoretical physics special
fund, NSFC.

\end{document}